# Will the "Butterfly Cipher" keep your Network Data secure?
## Developments in Computer Encryption

Vita Hinze-Hoare
Southampton University

**Introduction**

Encryption has been used as a method of protecting computer privacy since 1977. Secret communication plays an increasing role in organizations especially banking, industry, commerce, telecommunication etc. The basic idea of encryption is to modify the network traffic in such a way that its content can be reconstructed only by a legal recipient. This has been central to the setting up of secure systems such as Virtual Private Networks by establishing encrypted data tunnels trough the public Internet. (A general Classification of Encryption systems is shown in Appendix A)

The three-fold goal of network security is to protect **confidentiality** maintain **integrity** and assure **availability**. (**David Argles, 2005, Cisco Systems 2004, p5, Dieter Gollman 2004, p5, Tanenbaum and van Steen, 2002, p415**).

Organizations have been subject to a variety of attacks revealing network weaknesses at every level. Each of the vulnerabilities of the system can be best seen in terms of their main weaknesses at the individual level in the OSI model as shown below. The vulnerabilities are further described in **Appendix B**.

| Layer | Function | Devices/Protocols | Vulnerability |
|---|---|---|---|
| 1 Physical | Binary transmission | Connectors, Media | Wire tapping, sniffing, vandalism, power failure, theft, natural disasters, sabotage, disaffected personnel |
| 2 Data Link | Media access | MAC, LLC | Reconnaissance, sniffing, spoofing, broadcast storms, misconfigured NICs, attack robots |
| 3 Network | Address | IP, IPX, | Ping scans, Packet sniffing, ARP poisoning, DDOS, Smurf, Stacheldraht, Ping of death, Nuking |
| 4 Transport | End to End connection | TCP, UDP | Port scans, spoofing, session hijacking, DOS, SYN flood, UDP bombs |
| 5 Session | Inter Host communication | NFS, RPC | Traffic monitoring, share vulnerabilities, root access |
| 6 Presentation | Data representation | ASCII, HTML, PICT, WAV | Intercepting unencrypted data, compressed Trojan and viruses |
| 7 Application | Network processes to applications | Telnet, FTP, OS, Unix, HTTP, RMON, DNS, WHOIS, Finger | Email bombs, spam, Trojans, viruses, brute force attacks, holes in OS, worms, browser holes, malicious Java, activeX, CGI exploits, mapping and reconnaissance, DNS killer, control daemons, access permissions, key loggers, redundant accounts |

Most network weaknesses have been shown to involve human error or malice **(Gary Schneider 2006)** rather than technological issues, and for this reason Security policies (**Cisco 2004 p158**) have risen to the fore in the operation of companies and agencies (Appendix C shows trends driving Network Security in Industry) in order to enforce basic standards such as rotating passwords and confidentiality etc. (See Appendix D for details of the content of modern security policies).

The increasing use of computer networks in organisations brings an increase in risk to the organisational data assets from substitution, modification, theft, corruption and destruction. (For recently discovered network vulnerabilities see Appendix E). Organisations need to arm themselves against attack, particularly on publicly exposed websites (Web security issues are listed in Appendix F) and consequential loss with effective security measures that counteract those threats and protect their system's vulnerabilities. Recent developments in security defence are shown in Appendices G and H.

**Recent Developments in Encryption**
Current developments in encryption methods have moved towards the realm of atomic physics and involve the use of Quantum cryptography. While public-key cryptography relies on the computational difficulty factorizing large integers, (Appendix I summarises digital signatures/ Public Keys) quantum cryptography relies on the laws of quantum mechanics. This provides a secure method for transmitting a one time pad (secure key) over the network using quantum mechanical principles first explained by Einstein, Poldalsky and Rosen (EPR) in 1939. This is based on the idea that two photons given off in a single interaction will always be oppositely polarised. The strange quirk of quantum mechanics adds to this that by measuring the polarisation state of any photon you change its orientation. By measuring the orientation of a linked photon the orientation of the other photon is correspondingly changed at speeds faster than the speed of light no matter how far apart they are. This provides the mechanism for securely transmitting information from **A** to **B** using the EPR principle. **(Ekert, A 1995).**

Information is encoded in a pair of photons in such a way that any effort to monitor them necessarily disturbs them by some detectable means. This arises because of Heisenberg's uncertainty principle which maintains that certain pairs of physical properties, such as a photon's polarization, (e.g. rectilinear (vertical and horizontal) and diagonal (at 45° and 135°), cannot be measured without disturbing the other. This happens no matter how far apart the particles may be at the time and thus enables long-distance quantum key distribution. In this way entangled photon states are used to encode the bits and the information defining the key only "comes into being" after measurements performed by Alice and Bob.

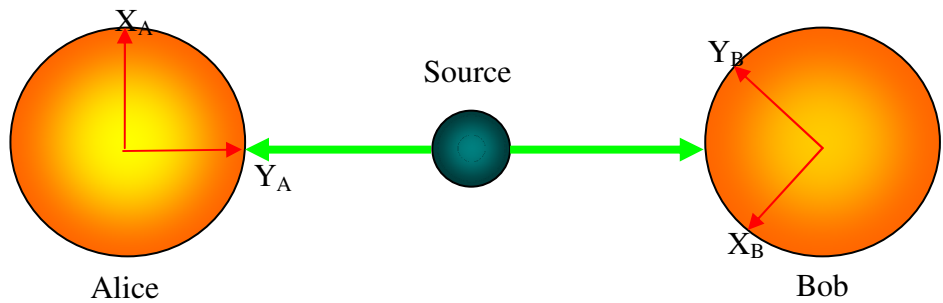

**Figure 1 Signalling using quantum cryptographic methods**

The advantage of this system is that in Quantum Cryptography, the traditional man-in-the-middle attack is impossible due to Heisenberg's uncertainty principle. If there is any interception of the stream of photons, they will be altered and the link will be destroyed. However this would not prevent a denial of service type attack by physically severing the connection. Despite this the communication is still secure and the essence of the message cannot be decoded. **(Ekert, A, 2005)**

This method is moving out of development into commercial practice. Quantum cryptography devices are already on the market from a few vendors (ID Quantique or MagiQ.) and looks like replacing such protocols as Diffie-Hellman key exchange in some areas. **Id Quantique (2005)** have a white paper describing "*how the deployment of quantum cryptography can help an organization to close loopholes in its information security architecture and reduce its vulnerability to cyber crime.*"

However the cost of the needed equipment, dedicated fiber optic line, the requirement to trust the equipment vendor, and the lack of a demonstrated threat to existing key exchange protocols all stacks against quantum cryptography at the moment.

**Encryption and chaos theory**
An even newer method of encryption published in Nature recently (Nature, Volume 438, Number 7066, pp298, 17 November 2005) shows that protecting data security and confidentiality need not involve complex technologies such as quantum encryption. Edward Lorenz's, 1972 chaos theory has been adapted to provide almost unbreakable encryption methods. Lorenz started his investigation into chaos theory by declaring that the "flapping of a butterfly's wing in Brazil could set up a tornado in Texas".

His mathematical representation of the unpredictability of systems has led to the idea of hiding messages in chaos. The idea is to encode data on top of a carrier wave composed of chaotic laser oscillations. These can be received by a system that matches the exact pattern of chaos used to encode the message. The process involves two matched lasers, the first to encode the signal message and the second identical laser to decode it. By using coordinated feedback, the laser is pushed into overdrive producing a chaotic mix of frequencies. The feasibility of this was first demonstrated in 1998 by van Wiggeren and Roy (Science, Vol 279, pp1198). More recently a team in Athens have shown that chaotically encrypted messages can be passed over standard telecom company cables up to a distance of a 120 km.

Figure 2 shows the message being added by the modulator to the chaotic light from the transmitting diode laser. This is unreadable until it is fed back into an identical diode laser producing the same chaotic signal. This is subtracted from the encryption by the splitter to reveal the message. ***Source: New Scientist, Vol 188 No. 2526, 19 Nov 2005, pp34***

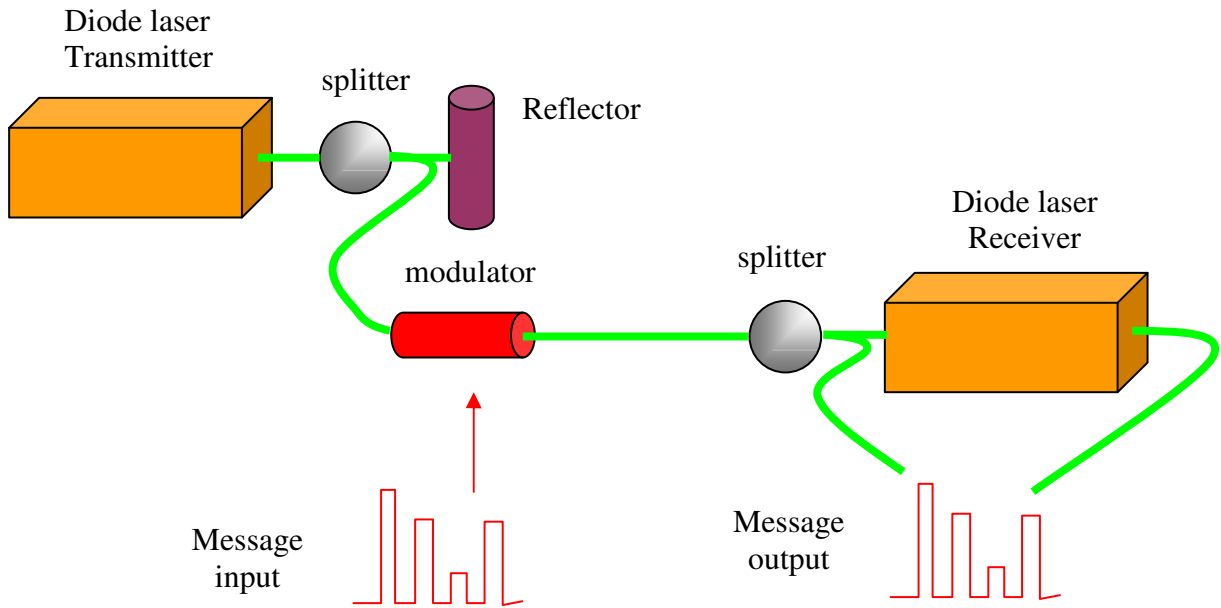

*Figure 2 Encryption and Chaos Theory*

**The idea of Discrete Chaotic Cryptography**
The theoretical basis behind chaos cryptography involves the application of non-linear mathematics which generates a transformation of the original text as follows:

First the plain text *C* is converted into a series of numbers; these are then transformed by a function φ that represents the chaos property of the dynamical laser system. Encryption is performed by repeating the transformation *n* times.

Cipher-text *P* is a result of the encryption:

$$P = \varphi^n(C) = \varphi(\ \varphi(\ ..\ \varphi(C)));$$

Decryption is performed by taking the inverse of φ a corresponding number of *n* times:

$$C = \varphi^{-n}(P) = \varphi^{-1}(\ \varphi^{-1}(\ ..\ \varphi^{-1}(P)));$$

*Zbigniew Kotulski, Janusz Szczepański, (1997)*

In order to intercept any message an eavesdropper would have to have exactly the right equipment. First they would need an optical coupler to tap into the chaotic light source. Next a duplicate of the of laser that constructed the message is needed. Finally they would need to find the same operating conditions and feedback control as the sender. Alternatively, if they tried to crack the code with brute force the high bit rates would require the transmission to be recorded to picosecond precision which is nearly impossible. In addition to that digitising the data stream would further distort the analogue signal creating further problems.

The implication for organizations is that almost perfect encryption can be used by telecom companies with a minimal change to the existing networks. Wide area networks and VPNs could be offered chaotic encryption to ensure confidentiality. The effect of this would be considerable. Banks could use it to back up their financial data and businesses could set up their own hacker proof private networks. In addition Government agencies and security services maybe the first to adopt this. However it is likely that the effects could be far reaching providing encryption for all telephone calls and cell phone signals as well as for sending secure signals to re-program satellites. The routine encryption of phone calls is something that civil liberty campaigners would welcome while government agencies would not.

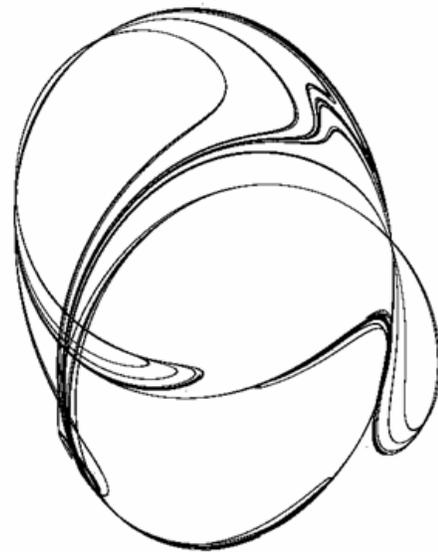

Compared with Quantum cryptography chaotic encryption is much faster. The former is a fairly slow process but chaotic cryptography can send data at around 2GBps which is typical of speeds required in standard optical telecommunications (New Scientist, vol.188, No256.pp34). However, quantum cryptography is inherently more secure because any attempts to listen in disrupt the delicate quantum states which will make the message unreadable. Chaotic cryptography does not have the same resistance to eavesdropping. Furthermore a recent paper by Kevin Short (Physical review letters, vol 83 pp 5389) has shown that chaotic cryptograms are all breakable in principle though very difficult to do in practice.

**Figure 3 Chaos Theory Generated Image Paul Bourke (1990)**

### Conclusion

*"Encryption does not protect against destruction of data or interruption of communication"*
**(D.Argles, 2005)**

A perfect encryption system will not solve all network/computer vulnerabilities. It will not prevent attacks on the physical layer where sabotage and natural disasters can occur. Nor will it prevent reconnaissance and broadcast storms at the data link layer. However encryption starts to become relevant at the network layer and above where the IP protocol can be made more secure against intrusion, packet sniffing etc. It would not protect against denial of service attacks. Where encryption will make the biggest difference is at session and presentation layers to enable the sharing of completely secure data streams. This will make the business to business connections over intranets and extranets more secure. An Organisation could offer chaotic encryption on dedicated networks to secure financial and commercially sensitive data. The implementation of quantum cryptography methods would further ensure that eavesdropping attempts would carry an instant alert thus preventing man in the middle attacks. However quantum cryptography is difficult to implement because the construction of the states is exceedingly difficult requiring an atom by atom preparation affecting any serious bandwidths use limiting data flow. This problem is not shared by the chaos encryption methods and ultimately this may prove to be the more practical of the two. However from an Organisation's point of view the data encryption is not so secure nor is intrusion detection guaranteed.

The goal of total security is one that would probably not be welcomed by a number of agencies. Top of this list would be Government agencies who monitor the flow of information not only within email but also around the internet in various ways. They have devoted a considerable number of resources to eavesdropping for such concerns as anti terrorism and defence. Should an unbreakable encryption algorithm be invented then this area would be the first to outlaw it. For years the US Government prevented the sale of encryption programs greater than 128bit to anywhere outside its shores. In addition the whole of the encryption software producing sector would be put out of business resulting in the loss of 12 billion dollars of revenue per year and high unemployment and retraining costs.

It may be questionable whether the main beneficiaries of a perfect encryption security system would be legitimate organisations or those on the darker side of society including terrorists, drug traffickers, espionage, organised crime etc. Some might therefore wonder whether perfect security is a goal worth seeking.

# APPENDIX A   General Classification of Encryption Systems

Encryptions systems can be classified either according to the encryption algorithm used or according to the method key distribution.

| Classification | Type | Description |
|---|---|---|
| *Structure of encryption algorithm.* | | |
| | **Stream ciphers** | A stream cipher system hides the plain-text by changing the bits of it in a random-like way. An interceptor, who does not know the key, will not know which bits have been changed. An ideal stream cipher would use a physical (true) random number generator as key generator but since its output cannot be reproduced, decipherment would be impossible. |
| | **Block ciphers** | Whole blocks of bits are treated simultaneously. A cipher-text block can be deciphered to any combination of plain-text bits or to as many combinations as there are keys. Commonly used block ciphers include DES, IDEA, FEAL, etc. |
| *Method of distribution of the key.* | | |
| | **Private secret key (Symmetric Key)** | Traditional cryptography requires that the sender and receiver of a message have the same secret key: the sender, to encrypt the message; the receiver to decrypt the cipher. This is known as secret key cryptography. The problem is passing the secret key from the sender to the receiver without anyone else finding out. |
| | | <table><tr><th>Type</th><th>Length of Key</th><th>Mechanism</th></tr><tr><td>Blowfish</td><td>1-448bits</td><td>Old and slow</td></tr><tr><td>DES Data Encryption Standard</td><td>56bits</td><td>Plain text encrypted in blocks. Too weak to use now</td></tr><tr><td>Triple DES</td><td>168bits</td><td>Same as above with two keys and three stages used</td></tr><tr><td>AES Advanced Encryption Standard</td><td>128 -256 bits</td><td>RIJNDAEL Standard adopted supports block sizes</td></tr><tr><td>Twofish</td><td>128-256 bits</td><td>Created by Bruce Schneier very strong</td></tr></table> |
| | **Public key (Asymmetric key)** | The public key cryptography was invented in 1976 by W.Diffie and M.Hellman. In this system, each person gets a pair of keys, called the public key and private key. Each person's public key is published while the private key is kept secret. All communications involve only public key, and no private key is even transmitted. *Example - RSA (Rivest, Shamir, Adelman, 1977)* |
| | | <table><tr><th>Type</th><th>Date</th><th>Mechanism</th></tr><tr><td>RSA (Rivest,Shamir,Adleman)</td><td>1978-2003</td><td>Requires keys at least 1024bits long for good security based on factoring large numbers into primes</td></tr><tr><td>KNAPSACK (Merkel&Hellman)</td><td>1978</td><td>Not effective</td></tr></table> |

**Classification with respect to the methods of constructing the algorithm.**

| Main tools applied in Cryptology | Methods |
|---|---|
| Number theory | Exploiting the difficulty in factoring large prime numbers e.g. RSA |
| Algebra | Algebraic transformations which are difficult to reverse |
| Algebraic geometry | Recently: elliptic curves over finite fields [7]) |
| Combinatorics | The use of sets and set theory to establish transformations. |
| Systems with large complexity | Chaos theory |
| Development of hardware and software | Hardware improvement makes cryptosystems more secure.  Zbigniew Kotulski, Janusz Szczepański, (1997), |
| Quantum Cryptography | Using the principles of quantum entanglement to establish the passage of a secure key |
| Chaos Cryptography | Using the principles of chaos theory to secure keys |

# APPENDIX B: SECURITY THREATS AND VULNERABILITIES INCLUDING:

| **Reconnaissance** | Unauthorised discovery and mapping of systems and services |
|---|---|
| **Eavesdropping** | Gathering Information on the network by packet sniffing |
| **Unauthorised System access** | Gaining access to a device for which the intruder does not have an account/password |
| **Man in the middle attack** | Hacker stands between two communication points gaining information |
| **Data manipulation** | Allows the intruder to capture ,manipulate and replay data (altering WebPages) |
| **IP spoofing** | Masquerading as someone else by falsifying the source IP address |
| **Session replay** | Capturing a sequence of packets and replaying them to cause unauthorised actions |
| **Auto Rooters** | Programs that automate the entire hacking process (Scan Probe, Capture) |
| **Back Door attacks** | Pathways into the system created by the hacker to allow easy re-entry |
| **Social engineering** | Trick a member of the Organisation to reveal their password (Pay Pal Accounts) |
| **Denial of Service attacks** | Attacker disables or corrupts the network by crashing the system or slowing it to a point where it is unusable |
| **Ping of death** | Modifying IP portion of the header to indicate more data in the packet than there is causing system to crash |
| **SYN flood** | Randomly opens up many TCP ports tying up the network with bogus requests thereby denying the service to others |
| **Packet fragmentation** | Exploits a buffer overrun bug in the internet working equipment |
| **Email bombs** | Send bulk emails to individuals thus monopolising their services |
| **CPU hogging** | Trojan horses or viruses tie up CPU cycles or memory |
| **Malicious applets** | Java, Java Script or ActiveX Programs that cause destruction to or tie up computer resources |
| **Reconfiguration of routers** | Rerouting traffic to disable web traffic |
| **Stacheldraht attack** | Automated tools used to remotely attack a single system using a large number of distributed systems together with encryption |
| **Tribe flood network** | Coordinated denial of service attack from many sources against one target |
| **Smurf attack** | Sends a large number of PING requests to broadcast addresses producing an avalanche of Pings which are returned to the spoofed IP address disabling the service |

**APPENDIX C: Recent trends driving security in Industry**

A number of trends are driving the demand for secure networks in this fast growing Industry. These include; wireless access, increased bandwidth, legal issues, privacy concerns, people shortages and increased threat to national security (terrorism). In the diagram below is shown the key issues which Companies have said most influenced them in setting security policies **(Cisco Systems 2004)**

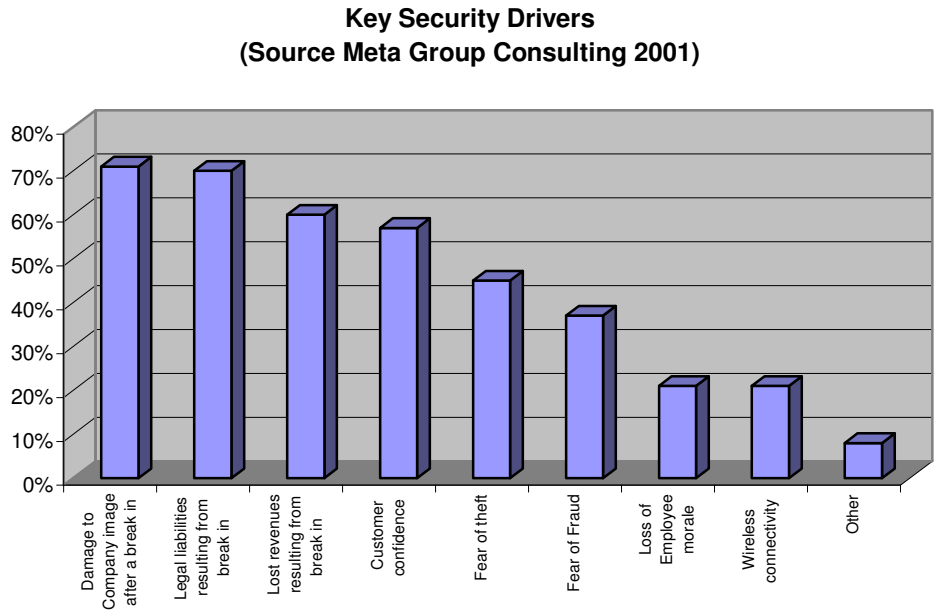

# APPENDIX D: IMPLEMENTING A SECURITY POLICY

**Security Policy**
Most security incidents occur because system administrators have not implemented a consistent security policy that benefits the organisation in a number of ways:

- Provides a process to audit the existing security
- Framework for implementing network security
- Defines behaviour which is and is not allowed
- Defines a process for handling security incidents
- Enables global security enforcement
- Creates a basis for legal action

The following table shows the important features of a security policy

| Feature | Description |
| --- | --- |
| Statement of Authority | Specifies who sponsors the security policy |
| Acceptable use policy | Specifies what the company will and will not accept regarding its information structure |
| Identification and Authentication policy | Specifies what technologies and equipment the company will use to ensure that only authorised people have access to its data |
| Internet access policy | Specifies what the company considers ethical in its use of Internet access |
| Site access policy | Specifies how on site users can use the company data |
| Remote access policy | Specifies how remote users will access the company data |
| Incident handling procedure | Specifies how the company will determine the procedures it will use during and after an incident – disaster recovery |

Cisco (2004) recommends the implementation of a continuous process of updating security policies called the security wheel, which involves the four tasks :( also illustrated overleaf)
- **Secure** – prevent unauthorised access and activities by applying the security policy and implementing authentication, firewalls, VPNs and vulnerability patching
- **Monitor** - active and passive methods of detecting security violations through system auditing and real time intrusion detection
- **Test** - security of a network is proactively tested using vulnerability scanning tools such as SATAN, NESSUS, and NMAP
- **Improve** – involves the analysing of the data collected during monitoring and testing and applying improvement mechanisms which then feed back into step one.

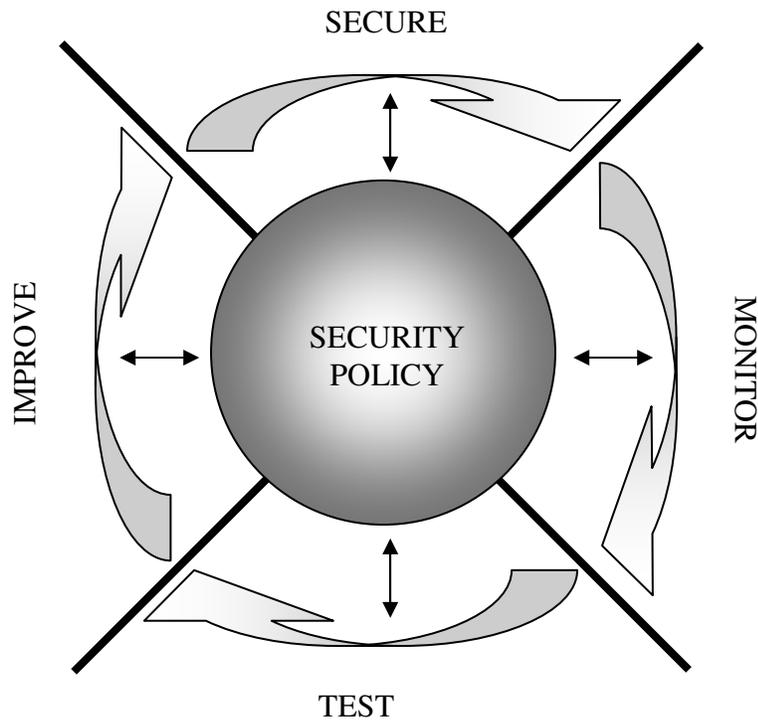

The key concept of establishing the identity of the user incorporates the triple AAA security e.g.
- **Authentication**, the process of validating the identity of an end user or device such as severs, switches, routers etc. (Identifies who the user is)
- **Authorisation** (Access Control), the process of granting access rights to a user or specified system and limiting the flow of information to only authorised persons or systems. (Defines what the user can do)
- **Accounting**, this is the process of logging each of the user activities (records what the user did).

**Main Authentication methods**

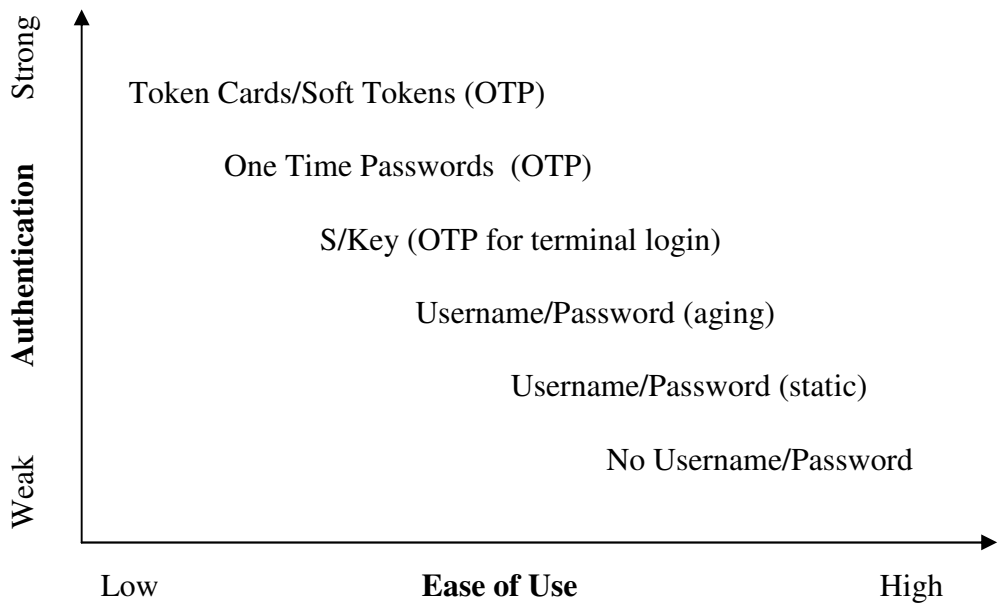

In practice the control of user identity for enterprises networks is performed by routers, switches, VPNs, firewalls, DSL, VOIP, network devices enabled by TACACS+ and RADIUS.

A number of authentication protocols have been established for use on insecure computer networks. These include;
- Authentication based on a share secret key
- Authentication based on Diffie Hellman key exchange
- -Using a key distribution centre
- - Kerberos
- -Public key cryptography

**Authorisation (Access Control)**
This involves "An active subject accessing a passive object with some specific access operation while a reference monitor grants or denies access" (Gollman p 31) This is the fundamental model of access control determined by Lampson (Bib82)

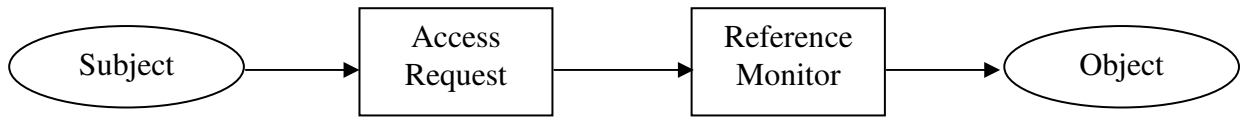

The subject can be any user or device while the object is typically a file or a resource. The reference monitor acts as a gateway policeman to ensure that only the right users have access to the appropriate resources. The most well known security device is the Firewall, which is a system that enforces an access control policy between two or more networks. These can range from dedicated appliances with integrated IOS (Internet Operating Systems) designed to provide enterprise level security for companies, to small personal software firewalls designed to protect the users PC such as zone labs, Norton.

Networks are secured through the use of access control lists (ACLs), which are used to filter and secure the network traffic. This is done by controlling whether routed or switched packets are forwarded or blocked. Each packet is examined by the router or switch and the decision whether to forward or drop the packet is based on the information supplied within the ACL.

**Accounting**
This function will keep log files that detail the various transactions of the user. The whole process of AAA is usually dedicated to a particular server. One such system is the CiscoWorks VPN/Security Management Solution which allows the configuring, monitoring and auditing of VPNs, Firewalls, IDS sensors via a Web Browser.
Accounting will also be concerned with an intrusion detection system, which audits host log files and notifies network Managers when an external process tries to modify a system file in a forbidden way. Cisco also provides an advanced form of IDS for the Cisco Security Agent at enterprise level.

**APPENDIX E: Recent Network vulnerabilities discovered**

**Skype Holes**
As VOIP continues to grow and become a dominant method of communication we can expect this to become the prime target for hackers. The ability to send spam telephone messages to millions of VOIP telephones at no cost might see an explosion of audible junk mail with a consequent need to employ call filtering and VOIP security. Skype, which has reached downloads of over 60m, (30% by businesses) has recently had two security holes identified. **(www.computerworld.com, 27$^{th}$ October 2005).** Both involve overflow errors which would enable attackers to execute any code in the system compromised or crash the system. Skype is particularly vulnerable to attack because it allows users to establish a direct connection and it is "port agile" e.g. if it finds a port is blocked it will search for an open one and as a result will make a connection 99 out of a 100 even behind a fire wall or a NAT *"As a result, Skype could provide a backdoor entry into otherwise secure networks for Trojans, worms and viruses."* ***Computerworld Oct. 2005***

**Cisco Router Holes**
Another recent vulnerability has been found in Cisco routers (**ZDNET 2005**) Cisco confirmed the existence of some serious vulnerabilities when Michael Lynn warned about when he demonstrated the ability to hack in to Cisco routers back in July. In response, Cisco issued this advisory warning all Cisco customers to upgrade their routers with the latest IOS. Cisco has also recently suffered from the publication by a group calling itself the Evil Scientists of a tool to reveal the saved passwords that reside inside the Cisco VPN Client. In addition they have also published exactly how the stored keys are encrypted. **(Loeb, L. 2005)**

**Recent Developments in Network Defence**
Network defence depends upon packet filtering and Access Control Lists (ACL). System administrators filter packets using ACLs on both inbound and outbound traffic. These are held on the routers at the perimeter of the network and reduce network threats while allowing legitimate traffic to pass. The very latest version of Cisco IOS software includes context based access control called CBAC which allows the setting of audit trails, global timeouts, port ton application mapping, defining and applying inspection rules, testing and verifying the packets.

| Type | Capabilities/Features |
|---|---|
| Standard ACL | Controls traffic by comparing source address of IP Packet to configured ACL address |
| Extended ACL | Compares source and destination packets |
| Lock-and-key ACL | Applies extended ACL, allows users to traverse router using telnet to enter key |
| IP-named ACL | Allows standard and extended ACLs to be given names instead of numbers |
| Reflexive ACL | Allows IP Packet filtering based upon upper layer session information |
| Commented IP ACL | Makes ACLs easier to understand |
| Context Based Access Control | Inspects traffic travelling trough firewall to discover state information that can be used to create temporary openings n the firewall's access lists |
| Turbo ACL | Processes ACL's more efficiently to improve performance |

# APPENDIX F WEB SECURITY

**Web Security (Tanenbaum)**

| Security Issue | Description | Defence |
|---|---|---|
| **DNS spoofing** | Breaking into DNS server and modifying its records enables the spoofer to pretend that his website is someone else's | Secure DNS can use random Ids in their queries. DNSsec is at present being researched and expected to take several years |
| **Interception** | **Sending security information across the Internet such as credit card number, needs to be kept secure from eavesdropping** | A secure sockets layer is interposed between the Application and Transport layers handling encryption. Converts HTTP into HTTPS |
| **Executable Code** | Small programs such as Java applets, activeX controls and Java script pose a substantial security risk if they contain malicious intent | **Java applets** run in an interpreter allowing every instruction to be examined before execution. **ActiveX** controls are binary programs embedded in web pages and these must be accompanies by digital signatures signed by creator before it can be run. **Java scrip**t can be switched off at the browser |

**APPENDIX G: Recent Developments in Security Defence**

**IDS/IPS**
Up until 2004 one of the most useful tools in the security armoury was the intrusion detection system IDS. However the limitation of this were clear insofar as it was only effective once an attack had already been made and been successful. Since that time the latest Generation of intrusion prevention firewalls IPS have come into their own. The Gartner report 2003, declaring the death of IDS although premature herald the way for such products as the Cisco Security Agent management system CSA **(Northcutt, 2005, p276).**

**Tripwire**
This is a file and directory integrity checker that accompanies Linux installation disks. **(Cisco 2004 p788 -792)** Its purpose is to check the files and directories don't contain errors and have not been damaged by a hacker. Tripwire works by creating a database of signatures for all files on the system. It compares new signatures with old ones on the regular basis to determine discrepancies exist. There are many ways to authenticate users in the Linux system the best being determined by the kind of server being used. For a remote access or telnet server secure remote passwords SRP can be used or open SSH. If it's a web of FTP server open SSL authentication should be used.

**IBM AXE**
Another recent development relates to IBM's Assured Execution Environment (AXE) project which is designed to police every program run on the system and denies access to everything which is not specifically authorised. This turns on its head the standard approach that the computer decides what needs to be run. AXE prohibits all code from running unless it's been preconfigured into a special format. This effectively gives each machine its own unique operating system determined by the user's need making it impossible for any viruses, worms of Trojans to run.

This approach has been patented by IBM as AXE runtime software which is loaded into the OS kernel upon boot-up.

> *"Users or administrators could use a variety of techniques, including encryption, to ensure that unauthorized software could not be run without their permission. They could also use Axe to make sure that certain programs were run only on specific machines or even use Axe techniques to make data unreadable, to keep Word or PowerPoint documents away from prying eyes." Computerworld Oct 2005*

**APPENDIX H**

**Biometrics**
A further recent development is the increasing use of biometric data to establish identity. This has involved a variety of methods including the scanning of fingerprints, irises, ears facial characteristics. Recent development also include writing pads to collect signatures that also detect the form and pressure of the writing, retinal scanners to read the pattern of blood vessels at the back of the eye, palm readers that scan the whole hand rather than a single fingerprint or read the pattern of veins on the back of the hand.(G.Schneider 2006) The German Federal Government has recently declared that from 2007 all passports will contain the subject's fingerprints for authentication. Sat 1 News 1$^{st}$ Nov. 2005-11-01. Issues here involve the reliability of the recognition processes which at present technology levels are not sufficient to guarantee a high enough success rate.

However not all see Biometrics as the best way forward. Peter Gutman (2204) of Auckland University has indicated that biometrics is fraught with a number of problems which still need to be addressed. His paper "Why Biometrics is not a Panacea" provides the technical background for the effectiveness of the technology that you'll never find in any vendor sales literature, and documents its less-than-stellar track record in the field.

It is never-the-less probable that Biometrics has a bright future if only we look far enough ahead. This is likely to involve the simultaneous measurement of a number of personal features "on the fly" as you walk through a portal, cross correlated to enable an ID fix with eventual absolute accuracy. When this is accomplished identity will be established and verified within a few seconds of entering any security sensitive or commercial premises allowing transparent cashless financial transactions to take place without the exchange of a signature or even a credit card and or ensuring the right person operates the right computer.

# APPENDIX I: Digital Signatures/ public keys

These provide three things:

1. Authentication of the customer
2. Non repudiation – no party can deny the transaction
3. Protection against fraud

| Type | Mechanism | Weakness |
|---|---|---|
| Symmetric key signature | Central Authority knows everything-everyone trusts | Central Authority reads all messages |
| Public key signature | Constructs digital signatures publicly without compromising security | Disclosure of Secret Key destroys non repudiation as anyone can use it |
| Message Digests | Decouples authentication from secrecy. Based on one-way hash function e.g. MD5 and SHA-1 | Birthday attack. Yuval(1979) |
| Digital Certificates | Attachment to email or program that verifies that the sender is who they claim to be, provides method of non-repudiation and encryption. Contains the following 6 elements<br>• Owners identifying information<br>• Owners Public Key<br>• Date of validity<br>• Certificate Serial No<br>• Certificate Issuer<br>• Issuer's Digital Signature<br>Issued by Companies such as Thawte and VeriSign | Does not attest to the quality of the software but only to the identity of the company |

Sources: Tanenbaum, Computer Networks, 4th Ed. pp 756-763
Schneider, Electronic Commerce p452